\documentclass[nofootinbib,preprint,aps,draft,pre,superscriptaddress,citeautoscript,floatfix]{revtex4-1}

\usepackage{graphics}
\usepackage{graphicx}
\usepackage{mathrsfs}
\usepackage{bm}
\usepackage{color}

\begin{document}

\title{Multi-Ring Deposition Pattern of Drying Droplets}
\author{Mengmeng Wu}
\affiliation{Center of Soft Matter Physics and its Applications, Beihang University, Beijing 100191, China}
\affiliation{School of Physics and Nuclear Energy Engineering, Beihang University, Beijing 100191, China}
\author{Xingkun Man}
\email{manxk@buaa.edu.cn}
\affiliation{Center of Soft Matter Physics and its Applications, Beihang University, Beijing 100191, China}
\affiliation{School of Physics and Nuclear Energy Engineering, Beihang University, Beijing 100191, China}
\author{Masao Doi}
\affiliation{Center of Soft Matter Physics and its Applications, Beihang University, Beijing 100191, China}
\affiliation{School of Physics and Nuclear Energy Engineering, Beihang University, Beijing 100191, China}

\begin{abstract}

We propose a theory for the multi-ring pattern of the deposits
that are formed when droplets of suspension are dried on a substrate.
Assuming a standard model for the stick-slip motion of the contact line,
we show that as droplets evaporate, many concentric rings of deposits are formed, but are taken over by a solid-circle pattern in the final stage of drying.
An analytical expression is given to indicate when ring-pattern changes to solid-circle pattern during an evaporation process. The results are in qualitative agreement with existing experiments, and
the other predictions on how the evaporation rate,
droplet radius and receding contact angle affect the pattern are all subject to experimental test.


\end{abstract}

\maketitle

\section*{Introduction} \label{sec.1}

When a droplet of suspension placed on a substrate is dried, it leaves various patterns of
deposits on the substrate. A well-known pattern is the ring-like deposit left when
a coffee droplet is dried on a plate. The deposition pattern has been studied
for various combination of solutes, solvents, and surfaces, and a variety of patterns
have been reported in the past~\cite{Adachi95,Chang08,Moffat09,Orejon11,Seo17,F18,Sujata18}.

In many situations, the deposits form concentric rings called multi-rings. This pattern has been
observed in evaporating films~\cite{Xia96}, liquid droplets~\cite{Deegan00,Shmu02,Bi12}, and confined solutions ~\cite{Lin05,Lin06,Lin12}. Other
types of patterns, spoke-like or eye-like patterns are also observed~\cite{Zhang14,Li15}. Understanding
the mechanism of such patterns are interesting not only from scientific view point, but also
from a view point of applications, in connection to surface patterning
of optical devices~\cite{Derby10}, biomacromolecular recognition~\cite{Zhang08} and disease detections~\cite{Yakhno05,Tran12}.

In this paper, we shall focus on the multi-ring pattern, and present a simple theory for the
origin of this pattern. Multi-ring has been explained by the stick-slip motion of the
contact line (CL). When the CL is pinned, a flow is created
from the droplet center to its edge to supply the liquid to the
edge. This flow convects solutes to the edge and deposits them near the CL.
As the droplet volume decreases by evaporation, the
contact angle decreases, and creates an inward unbalanced force
(depinning force) acting on the contact line. When the contact angle
becomes less than the receding contact angle $\theta_{\rm{R}}$ (the
angle at which the contact line starts to recede),
the contact line starts to slip and move quickly towards the
center until it becomes pinned again. The repetition of this stick-slip
motion of the CL generates the multi-ring pattern.

Although the multi-ring formation has been explained qualitatively
by this mechanism, theoretical modeling for the process has been
undeveloped as the problem involves the fluid flow and the contact line
motion in evaporating droplets coupled with the particle
transport. Previous theoretical models ~\cite{Nono03,Frastia11,Zigel16}
are written in the form of non-linear partial  differential equations, and
required numerical simulation to see the outcome of the model.

It has been reported that the multi-ring pattern is usually
made of a solid circle in the center surrounded by many concentric rings, as schematically
shown in Figure~1. This phenomenon has indeed
been observed experimentally in drying of droplets containing a wide type of
solutes, including colloid~\cite{Adachi95,Shmu02,Yang14}, polymer~\cite{Bi12}, DNA~\cite{Chang08}
and nanoparticles~\cite{Jing16}. The open question that why and how a solid-circle often appears in
the central region of concentric rings has not been answered as far as we know.

\begin{figure*}[h!t]
\begin{center}
{\includegraphics[bb=0 0 582 297, scale=0.5,draft=false]{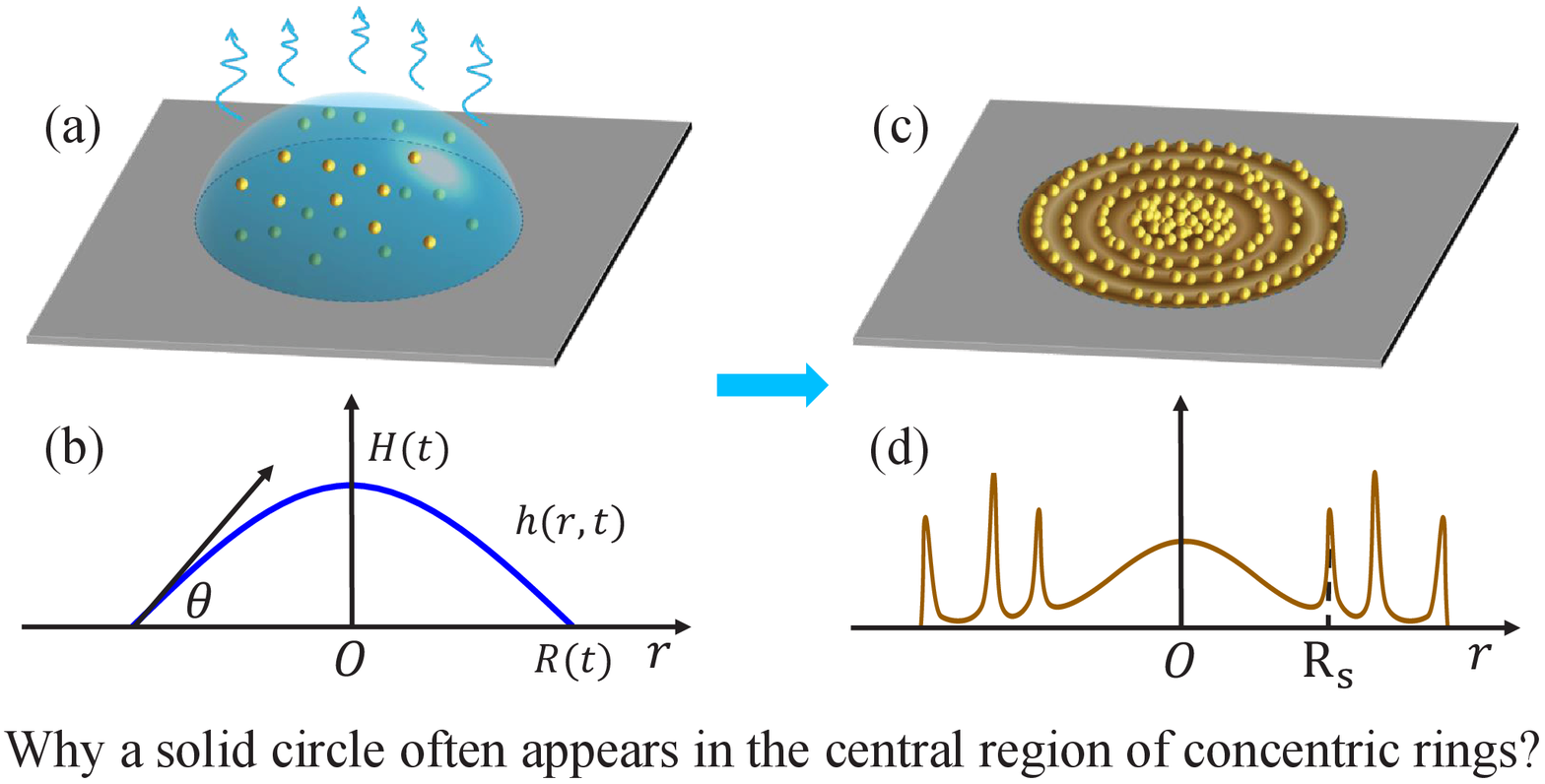}}
\caption{
\textsf{Schematic of the multi-ring pattern. (a) A droplet of particle suspension is dried on a substrate. (b) Relevant parameters are the radius of the contact line $R$, the hight of the droplet at the center $H$, and the contact angle $\theta$. (c) The top view and (d) the side view of the deposition pattern. The deposit is made of a solid circle in the center of concentric rings. $R_{\rm S}$ is the radius of the innermost ring.}}
\end{center}
\end{figure*}

In this paper, we propose a simple model for the formation process of
the multi-ring. This paper is an extension of our previous
work~\cite{Man16,Hu17} on the
contact line motion of an evaporating droplet. Using Onsager principle,
we derived first-order ordinary differential equations for the droplet radius $R$
and the contact angle $\theta$, and have shown that the deposit pattern
can change from ring-like (where the peak is a ring at the edge)
to mountain like (where the peak is located at the center). To explain
the change, we introduced a phenomenological parameter $k_{\rm{cl}}$
which represents the friction of the contact line, and is
assumed to be a constant determined by the interaction
between the liquid and substrate. In this paper, we assume that
$k_{\rm{cl}}$ is not constant and changes with the contact angle. We will use the simplest possible
model for $k_{\rm{cl}}$ to describe the stick-slip motion, and show
that this model captures the characteristic features of the multi-ring pattern, and also
can answer to the open question mentioned above. We also predict the condition for the multi-ring pattern to be observed, and address the key factors that determine the inter-ring spacing of the deposits.

\section*{Theoretical Framework} \label {sec.2}

\subsection*{Evolution Equations for Drying Droplets} \label{sec.2.1}

We consider an evaporating droplet containing nonvolatile solutes,
which is placed on a substrate. We assume that the droplet contact angle
$\theta(t)$ is small ($\theta < 1$), and the surface profile of the droplet is given by
(in a cylindrical coordinate)
\begin{equation} \label{eqn:fh}
  h(r,t)=H(t)\left[1-\frac{r^2}{R^2(t)}\right],
\end{equation}
where $H(t)$ and $R(t)$ are the height and the radius of the droplet.
The droplet volume $V(t)$ is then given by
\begin{equation} \label{eqn:vol}
       V(t)=\frac{\pi}{4} \theta R^3(t).
\end{equation}
Let $R_0$, $\theta_0$
and $V_0$ be the initial value of these parameters.

The droplet volume decreases in time by solvent evaporation.
The rate $\dot V$ is essentially determined by the diffusion of
solvent vapor in air. When there is no air flow near the liquid surface, $\dot{V}$ can be calculated by solving the diffusion equation of solvent vapor in air. Such studies~\cite{Parisse97,Kobaya10} have shown that $\dot V$ is proportional to $R(t)$.
Hence, we assume
\begin{equation} \label{eqn:dvol}
   \dot V(t)=\dot V_0 \frac{R(t)}{R_0},
\end{equation}
where the initial rate $\dot V_0 (<0) $ is a constant determined by
the initial droplet radius $R_0$, the temperature and the humidity of the environment.
The evaporation rate (the volume of solvent evaporating per unit
time per unit surface area) is given by
\begin{equation} \label{eqn:eva}
J(t)=-\frac{\dot V(t)}{\pi R^2(t)}=-\frac{\dot{V_0}}{\pi R_0R(t)}.
\end{equation}
This indicates that the evaporation rate increases as droplet
size $R(t)$ becomes small.

Given the volume evolution equation, we need one more equation either for $R(t)$ or $\theta(t)$ to describe the droplet shape evolution during evaporation. We determine the time evolution of these parameters by the Onsager principle \cite{Doi13,Doi15}.

This principle is equivalent to the variational principle known in Stokesian hydrodynamics which states that the evolution of the
system is determined by the minimum of Rayleighian defined by
\begin{equation}\label{fr}
\Re=\Phi+\dot{F}
\end{equation}
where $\Phi$ is the energy dissipation function (the half of the energy dissipation rate created in the fluids by the boundary motion), and $\dot F$ is the time derivative of the free energy of the system.
The principle has been applied for the droplet motion by gravity \cite{Xu16} and by evaporation \cite{Man16,Man17}.

We assume that the droplet is nearly flat [$R(t)\gg H(t)$] and use the lubrication approximation to calculate the dissipation function, which is written in the following form~\cite{Man16}
\begin{equation}\label{fphi}
\Phi=\frac{3\pi^2\eta R^4}{4V} \left[\ln\left(\frac{R}{2\epsilon}\right)-1\right]
                                              \left(\dot{R}-\frac{R\dot{V}}{4V}\right)^2
       +\pi\xi_{\rm{cl}} R\dot{R}^2
\end{equation}
where $\eta$ is the viscosity of the fluid, $\xi_{\rm{cl}}$ is a phenomenological parameter representing the mobility of the contact line, and $\epsilon$ being the molecular cut-off length which is introduced to remove the divergence in the energy dissipation at the contact line. The first term in Eq. (\ref{fphi}) represents the usual hydrodynamic energy dissipation in the lubrication approximation, while the second term represents the extra energy dissipation associated with the contact line motion over substrate.

The free energy $F$ is a sum of the interfacial energy by assuming the droplet size is less than the capillary length. The time derivative of such free energy is~\cite{Man16}
\begin{equation}\label{fdotf}
\dot{F}=\gamma_{LV}
      \left[ \left( -\frac{16V^2}{\pi R^5}+\pi\theta^2_e R \right ) \dot R
                +\frac{8V\dot{V}}{\pi R^4}
      \right].
\end{equation}
where $\gamma_{LV}$ is the liquid/vapor surface tension and $\theta_e$ is the equilibrium contact angle.

By minimizing the Rayleighian $\Re=\Phi+\dot{F}$ with respect to $\dot{R}$, we obtain the force balance equation for the contact line as
\begin{equation}\label{eqn:force}
\xi_{\rm{hydro}}\left(\dot{R}-\frac{R\dot{V}}{4V}\right)+
     \xi_{\rm{cl}}\dot{R}
         =  \frac{ \gamma_{LV}}{2} \left( \theta^2 - \theta^2_e  \right).
\end{equation}
The right hand side represents the unbalanced capillary force acting on the contact line,
where $\gamma_{LV}$ is the surface tension, and $\theta_e$ is the
equilibrium contact angle. The left hand side represents the frictional
force acting on the moving contact line, where $\xi_{\rm{hydro}}$
is the friction constant calculated by hydrodynamics~\cite{Bonn09}
and $\xi_{\rm{cl}}$ is the phenomenological parameter introduced to
account for the contact line hysteresis. The hydrodynamic
friction constant is expressed by the fluid viscosity $\eta$ as
$\xi_{\rm{hydro}}=3C \eta/\theta$, where $C=\ln(R/2\epsilon)-1$
is a constant arising from the molecular length scale $\epsilon$.

Equation (\ref{eqn:force}) can be rewritten as the time evolution equation of the droplet contact radius
\begin{equation}\label{eqn:drad}
\left(1+k_{\rm{cl}}\right)\dot{R}
         =  \frac{R\dot{V}}{4V}
             + \frac{\gamma_{LV}\theta\left(\theta^2-\theta^2_e\right)}{6 C \eta },
\end{equation}
where $k_{\rm{cl}}$ is defined by $k_{\rm{cl}}=\xi_{\rm{cl}}/\xi_{\rm{hydro}}$ and represents the importance of the extra
friction constant $\xi_{\rm{cl}}$ of the contact line
relative to the normal hydrodynamic friction
$\xi_{\rm{hydro}}$.
We use $k_{\rm{cl}}$ as a phenomenological parameter to
distinguish the stick state (where $k_{\rm{cl}} \to \infty$),
and the slip state (where $k_{\rm{cl}} \to 0$).

In the previous paper \cite{Man16}, we take
$k_{\rm{cl}}$ as constant, but now we assume that it has two
values, low value in the slip state and high value in the stick state,
which is written as
\begin{equation}\label{eqn:kcl}
k_{\rm{cl}}=\left\{
\begin{array}
{r@{\quad\, \quad}l}
     0 & \text{for} \
                         \theta \le \theta_{\rm{R}} \ \
                                \text{or}  \ \
                         \dot{\theta} > 0 ,\\
      \alpha & \text{for} \
                         \theta > \theta_{\rm{R}} \
                                \text{and} \
                         \dot{\theta} \le 0 ,
\end{array}
\right.
\end{equation}
where $\theta_{\rm{R}}$ is the receding contact angle (the angle below which the
contact line starts to recede) and $\alpha$ is a constant representing the CL moving ability.
We take $\alpha=100$ for all calculations, which is large
enough to make the CL stick. The contact line sticks
when $k_{\rm{cl}}=\alpha$, while it slips when $k_{\rm{cl}}=0$.

When $R$ is pinned, $\theta$ decreases
due to the evaporation. The deviation between $\theta(t)$ and
$\theta_e$ generates a capillary force to pull the CL inwardly, but
the CL remains stick as far as $\theta$ is larger than $\theta_{\rm{R}}$.
When $\theta$ becomes smaller than $\theta_{\rm{R}}$, the CL starts to recede,
and $\theta$ starts to increase. When $\theta$ becomes equal to
its maximum value, $\dot \theta$ becomes equal to zero, and the contact line starts to be pinned.
The repetition of this dynamics generates the stick-slip motion of the contact line.

To characterize the evaporation rate, we introduce
another dimensionless parameter $k_{\rm{ev}}$
by $k_{\rm{ev}}=\tau_{\rm{re}}/\tau_{\rm{ev}}$.
Here, $\tau_{\rm{ev}}$ is the evaporation time defined by
$\tau_{\rm{ev}}=V_0/|\dot{V}_0|$, and $\tau_{\rm{re}}$ is
the relaxation time defined by
$\tau_{\rm{re}}=\eta V^{\frac{1}{3}}_0/\gamma_{LV}\theta^{3}_e$.
If $k_{\rm{ev}}$ is large, the relaxation time is much longer than the evaporation time,
leading to that the droplet volume decreases
much faster than the equilibration of the contact angle,
therefore, $\theta$ becomes much smaller than
$\theta_e$. On the other hand, if $k_{\rm{ev}}$ is
small, $\theta$ remains close to $\theta_e$.

Equations (\ref{eqn:vol}), (\ref{eqn:dvol}), (\ref{eqn:drad}), and (\ref{eqn:kcl})
determine the time evolution of $V(t)$, $R(t)$ and $\theta(t)$.
These equations can be solved for given values of
dimensionless evaporation rate $k_{\rm{ev}}$,
and two contact angles $\theta_{\rm{R}}$ and $\theta_e$.

\subsection*{Deposition Mechanism}
The distribution of the deposits left on the substrate can be calculated by using the time evolution equations of the droplet volume and contact radius. Since the diffusion of the solute in radial direction can be ignored for macroscopic droplets \cite{Anderson95}, we can assume that the solute moves with the same velocity as the fluid as long as the solute is in the droplet. The height-averaged fluid velocity at position $r$ and time $t$, $v(r,t)$, is obtained by solving the conservation equation $\dot h =- \nabla\cdot(v h ) - J $, resulting in a simple expression as
\begin{equation}\label{fv}
v(r,t)=r \left(\frac{\dot{R}}{R}-\frac{\dot{V}}{4V}\right).
\end{equation}
A fluid flow from center to edge is induced when $v>0$. On the other hand, $v<0$ indicates a fluid flow from the edge to droplet center. Consider the particles located at $r_0$ at time $t=0$. Let $\tilde r(r_0,t)$ be the average position
of the particles at time $t$. Since the particles are carried by the fluid flow with the velocity $v(r,t)$ given by Eq.~(\ref{fv}), $\tilde r(r_0,t)$ satisfies the equation
\begin{equation}
\frac{\partial \tilde r}{\partial t}=\tilde r \left(\frac{\dot{R}}{R}-\frac{\dot{V}}{4V}\right)
\end{equation}
At some time, the particles arrive at the contact line and are deposited there. This happens at time $t_{d}$ when $\tilde r(r_0,t_d)= R(t_d)$. The amount of particles deposit at this point is then calculated by $\tilde r(r_0,t_d)$
as follows.

The total amount of solute which was originally contained in the region between $r_0$ and
$r_0+dr_0$ at time $t=0$ is $2 \pi r_0 h(r_0, 0) \phi_0 dr_0$. When this interval meets the CL, all solutes within this interval deposit in the region between $\tilde r$ and $\tilde r+d\tilde r$. Therefore, the density of deposit at the position $\tilde r$ is given by
\begin{equation}\label{fdeposite}
    \mu=h(r_0, 0) \phi_0 \frac{r_0}{\tilde r} \left( \frac{d \tilde r}{d r_0} \right)^{-1}.
\end{equation}
Notice that both the defined deposition time $t_d$ and position $\tilde r$ are functions of $r_0$. More details of the model and deposition mechanism can be found in the previous work \cite{Hu17}.

\begin{figure}[h!]
\begin{center}
\includegraphics[bb=0 0 335 266, scale=0.52,draft=false]{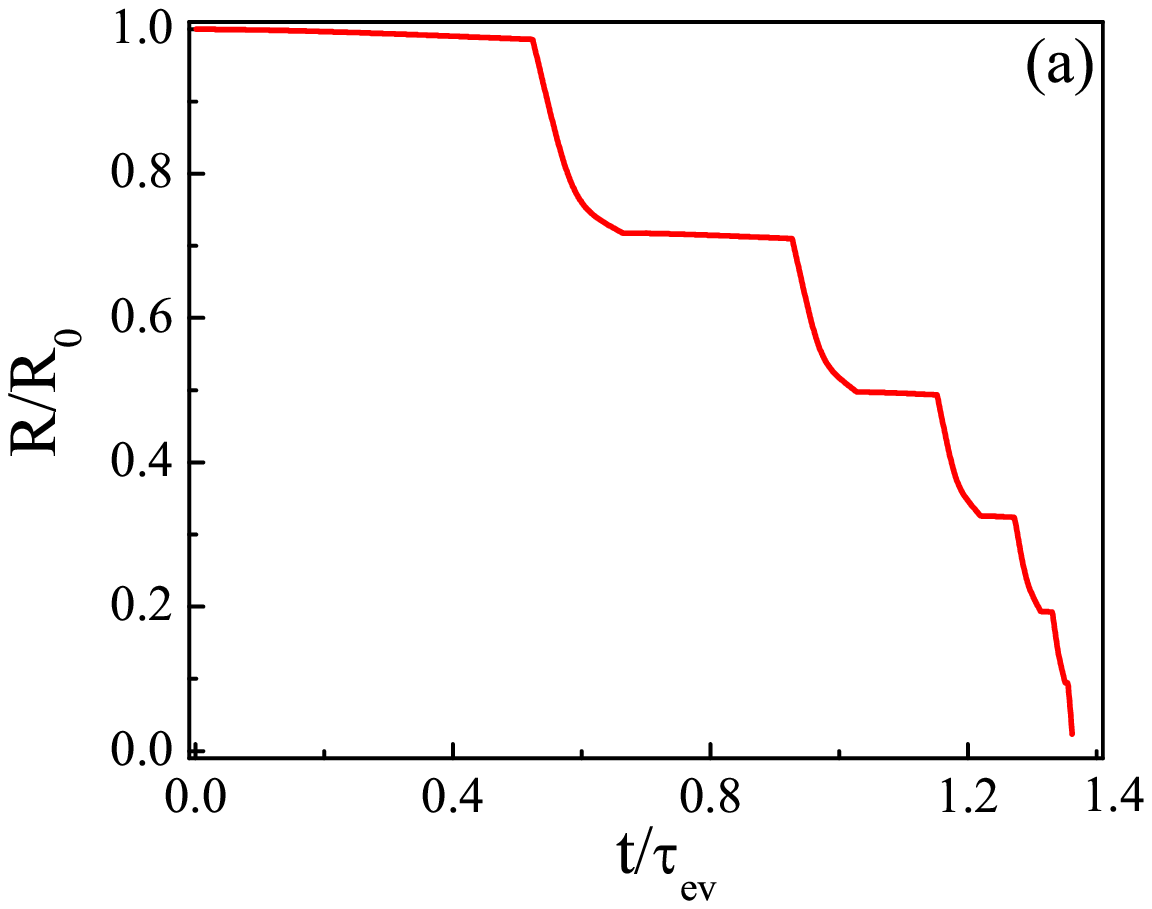}
\includegraphics[bb=0 0 335 266, scale=0.52,draft=false]{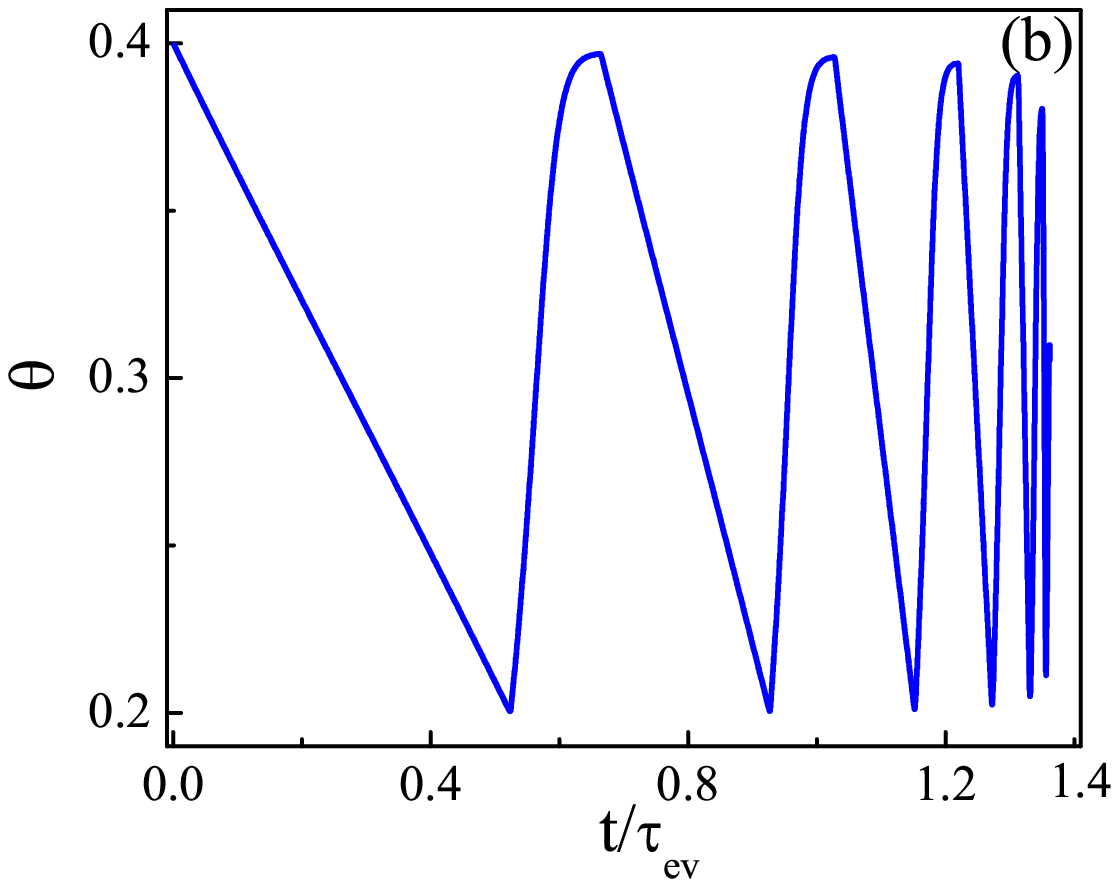}
\caption{Evolution of (a) the droplet contact line, $R(t)/R_0$, and (b) the droplet contact angle, $\theta$, during an evaporation process.
Here, the time is in units of $\tau_{\rm{ev}}$. The stick-slip motion of $R$ is observed, while the contact angle oscillates
between $\theta_e$ and $\theta_{\rm{R}}$. In both figures, the parameters are $k_{\rm{ev}}=10^{-3}$, $\theta_{\rm{R}}=0.2$, and $\theta_0=\theta_e=0.4$.}
\label{fig:motion}
\end{center}
\end{figure}
%

\section*{Results and discussion} \label {sec.3}

\subsection*{The Stick-Slip Motion and The Inter-Ring Spacing} \label {sec.3.1}

Typical droplet shape evolutions calculated from the model are shown in Figure~\ref{fig:motion}. Figure~\ref{fig:motion}a shows that $R(t)$ decreases stepwise, while Figure~\ref{fig:motion}b
shows that $\theta$ oscillates between the equilibrium contact angle
$\theta_e=0.4$ and the receding contact angle $\theta_{\rm{R}}=0.2$.
The stick period
becomes shorter as droplet size gets smaller since the evaporation rate increases as $R(t)$
decreases. In the final stage, the stick period gets so short that the CL looks to move continuously.

\begin{figure}[h!]
\begin{center}
\includegraphics[bb=0 0 230 180, scale=1.1,draft=false]{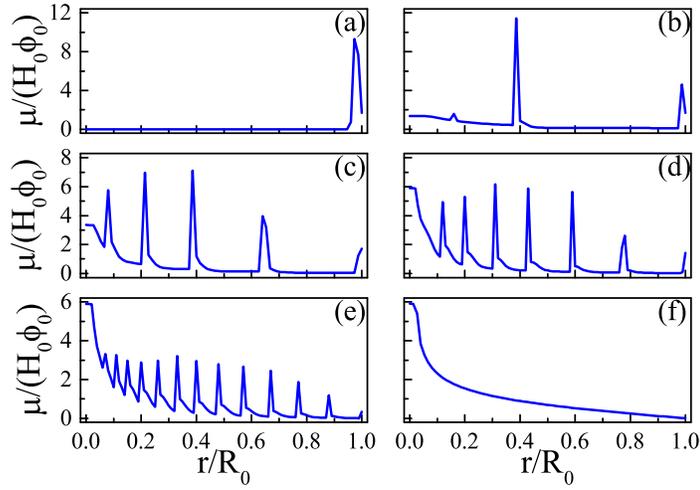}
\caption{
Different profile of the deposits left on the substrate when the drying
is completed for droplets with different values of $\theta_{\rm{R}}$, (a) $\theta_{\rm{R}}=0.00$; (b) $\theta_{\rm{R}}=0.08$;
(c) $\theta_{\rm{R}}=0.16$; (d) $\theta_{\rm{R}}=0.24$; (e) $\theta_{\rm{R}}=0.32$; (f) $\theta_{\rm{R}}=0.4$.
For all calculations, $\theta_0=\theta_e=0.4$, $R_0=4$, and $k_{\rm{ev}}=10^{-3}$.}
\label{fig:pattern}
\end{center}
\end{figure}

If the time evolution of the CL radius $R(t)$ is given, the profile of the deposit
can be calculated by the method described in the subsection of deposition mechanism.
Figure~\ref{fig:pattern} shows how the deposition pattern changes when the receding
contact angle $\theta_{\rm{R}}$ is changed.
Here the evaporation is assumed to be slow ($k_{\rm{ev}}=10^{-3}$),
and $\theta_0=\theta_e=0.4$.
It is seen that deposition pattern changes from coffee ring to multi-ring
and to mountainlike by increasing $\theta_{\rm{R}}$
from $0$ to $0.4$. When $\theta_{\rm{R}}=0$ (Fig.~\ref{fig:pattern}a),
the CL remains stick since $\theta(t)$ stays larger than $\theta_{\rm{R}}$
during evaporation, and coffee ring pattern is observed.
When $\theta_{\rm{R}}=\theta_e$ (Fig.~\ref{fig:pattern}f), the CL recedes freely, and
creates the mountainlike pattern. Between these limits (Figs. \ref{fig:pattern}b$-$\ref{fig:pattern}e),
various multi-ring patterns are obtained. As $\theta_{\rm{R}}$ increases,
the number of rings increases and the inter-ring spacing $\Delta L$ decreases
since the CL becomes easily depinned with the increase of $\theta_{\rm{R}}$.
Figure~\ref{fig:pattern} indicates that coffee ring and mountain-like patterns are
special cases of multi-rings: they can be regarded as a multi-ring
with $\Delta L \to \infty$ and $\Delta L \to 0$, respectively.

\begin{figure}[h!]
\centering
\includegraphics[bb=0 0 335 266, scale=0.6,draft=false]{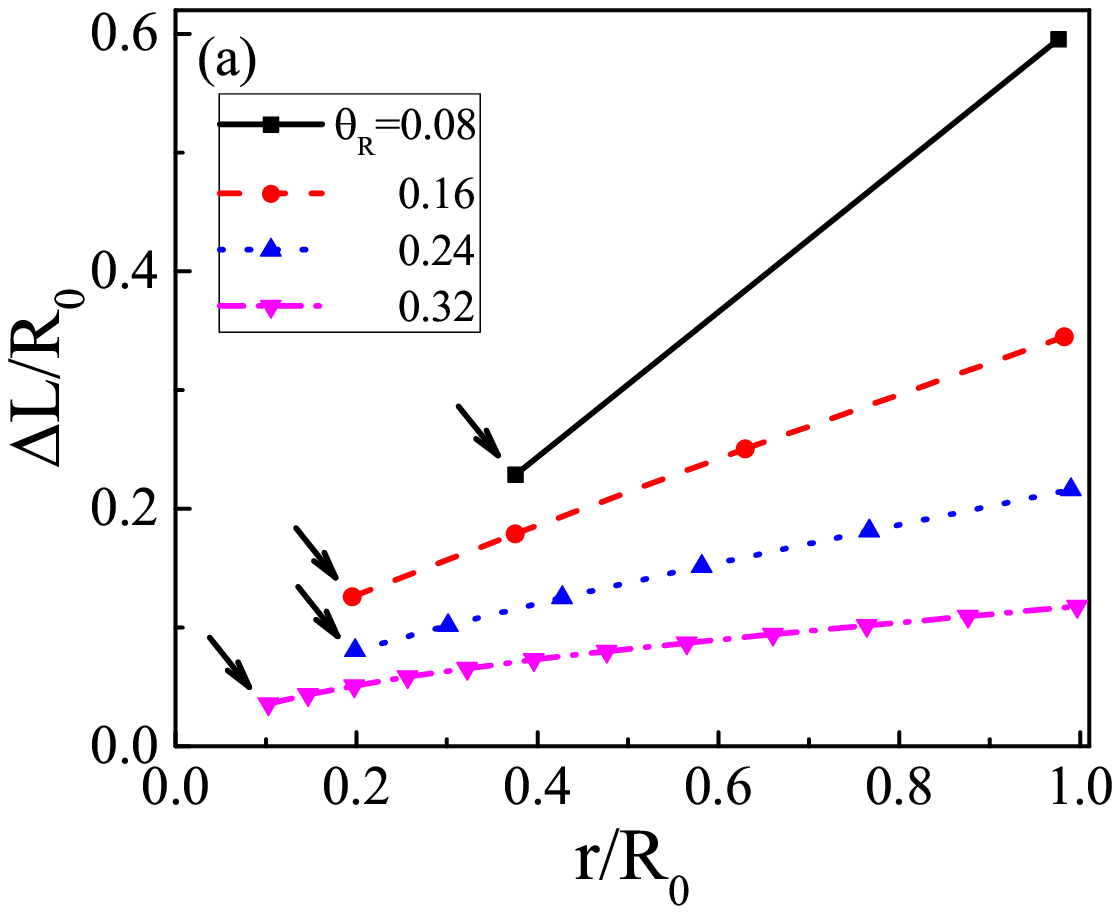}
\includegraphics[bb=0 0 335 266, scale=0.6,draft=false]{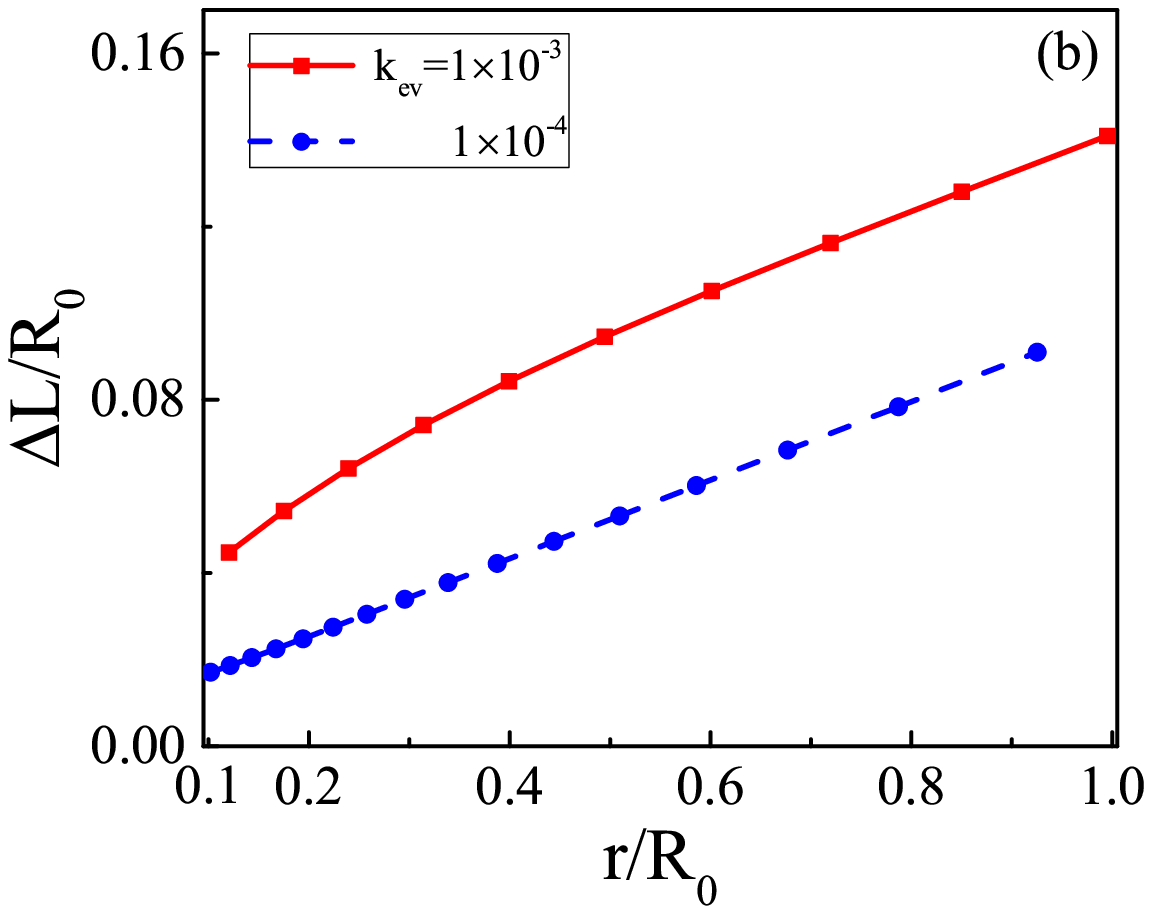}
\caption{The rescaled inter-ring spacing $\Delta L/R_0$ is plotted as a
function of the distance to the droplet center $r/R_0$ for different $\theta_{\rm R}$ and $k_{\rm ev}$. (a) As the
receding contact angle increases (or the contact angle hysteresis decreases),
the total number of rings increases, while $\Delta L$
decreases. The vertical arrows indicate the radius of the last ring of the
multi-ring pattern. The parameter $k_{\rm{ev}}=10^{-3}$. (b) As the evaporation rate ($k_{\rm ev}$) increases, $\Delta L$ increases while the number of rings decreases. The parameter $\theta_R=0.3$. For all cases, $\theta_e=\theta_0=0.4$.}
\label{fig:interring}
\end{figure}

Figure~\ref{fig:interring} shows the inter-ring spacing $\Delta L$ as a function of the rescaled distance to the droplet center $r/R_0$ for different $\theta_R$ and $k_{\rm ev}$. It is clear seen that $\Delta L$ is an increasing function of $r/R_0$ for all cases, indicating that rings are denser near the center than the edge. Figure~4a also shows that as $\theta_R$ increases, $\Delta L$ decreases and the number of rings increases, while Fig.~4b shows that as the evaporation rate $k_{\rm ev}$ increases, $\Delta L$ increases and the number of rings decreases.

Such tendencies have indeed been observed experimentally~\cite{Yang14,Lin06}.
Yang et al.~\cite{Yang14} measured
the distance between two successive rings and reported
that $\Delta L$ decreases almost linearly with the radius
$\Delta L \sim r$. The increase of $\Delta L$ as a function of
the ring radius is also observed in different experimental set-ups.
Xu et al.~\cite{Lin06} studied the multi-ring pattern of polymer solutions
evaporating between a sphere and a flat substrate. They showed that
the center-to-center distance between adjacent rings is an increasing
function of the distance from the touching point of the sphere on the substrate.
In both cases, the inter-ring spacing varies with the ring radius, and is
consistent with our theory.

The effects of $\theta_R$ and evaporation rate on $\Delta L$ are also consistent with experiments. Yang et al~\cite{Yang14} showed that $\Delta L$ is larger for a larger particle volume fraction. They explained that this is due to the increase of pinning force with the increase of particle concentration. The pinning force is related to the receding contact angle $\theta_R$ in our model: the larger the pinning force is, the smaller the receding contact angle is. Figure~4a shows that $\Delta L$ indeed increases when $\theta_R$ decreases from $0.32$ to $0.08$, which is qualitatively consistent with the experimental results. On the other hand, Xu et al~\cite{Lin06} showed that as the evaporation rate increases, the spacing $\Delta L$ increases, and the number of ring decreases. Figure~4b confirmed this experimental finding, where each dot in this figure indicates one ring of the deposition pattern.

$\Delta L$ can be estimated by the following simple argument. The CL starts to recede
when $\theta$ becomes equal to $\theta_{\rm{R}}$ and quickly moves
to the next pinning point where $\dot \theta$ becomes equal to 0 (or $\theta$ becomes close to $\theta_{e}$).
Hence the volume conservation equation is written as
\begin{equation}\label{eqn:deltL}
 \frac{\pi}{4} \theta_{\rm{R}} R^3
     = \frac{\pi}{4} \theta_{e} (R- \Delta L)^3 + \Delta V.
\end{equation}
where $\Delta V$ is the volume change of the droplet during the process.
For slow evaporation, we may assume $\Delta V \ll V$, then Eq.~(\ref{eqn:deltL})
gives
\begin{equation}\label{eqn:SdeltL}
\Delta L=R\left[1-\left(
                               \frac{\theta_{\rm{R}}}{\theta_{\rm{e}}}
                           \right)^{\frac{1}{3}}
                 \right].
\end{equation}
Equation~(\ref{eqn:SdeltL}) indicates that $\Delta L$ decreases as $R$ decreases,
but increases as $\theta_{\rm{R}}$ decreases, which qualitatively agrees with the numerical results in Fig.~\ref{fig:interring}a.
It is interesting to note that Eq.~(\ref{eqn:SdeltL}) explains the formation of
mountain-like and coffee ring patterns (since $\Delta L=0$
for $\theta_{\rm{R}}=\theta_e$  and  $\Delta L=R_0$ for
$\theta_{\rm{R}}=0$).

\subsection*{The Condition for Multi-Ring Formation} \label{sec.3.2}

In the previous subsection, we have shown that for the multi-ring
pattern to be observed, the contact line has to do stick-slip motion
(i.e., switching between pinned state and depinned state).
A condition for
this to happen is that the receding contact angle $\theta_{{\rm R}}$
is non-zero, but
this is not the only condition for the stick-slip motion.
In fact, the evaporation rate $k_{\rm{ev}}$
is another important factor, and its combined
effects with $\theta_{{\rm R}}$ in
the condition for multi-ring formation can be estimated theoretically.

First we rewrite Eq.~(\ref{eqn:drad}) for $\dot R$ to an
equation for $\dot \theta$:
\begin{equation}\label{eqn:dangl}
\left(1+k_{\rm{cl}}\right)\tau_{\rm{re}} \dot{\theta}
     = -\frac{ k_{\rm{ev}}\left(1+4k_{\rm{cl}}\right) V_{0}}{\pi R_0 R^2}
             + \frac{V^{\frac{1}{3}}_{0}\theta^2 \left(\theta_e^2-\theta^2\right)}
                            {2C \theta^3_{e} R
                             }.
\end{equation}
The first term on the right hand side is negative, while the second term is positive
since $\theta$ is less than $\theta_e$ during the evaporation.

Now consider a droplet in the stick state. As solvent evaporates,
the contact angle $\theta$ decreases (because $k_{\rm{cl}}$ is
large in the stick state). When $\theta$ becomes equal to
$\theta_{\rm{R}}$, $k_{\rm{cl}}$ switches from large positive
value to 0. In the usual stick-slip motion, this jump of
$k_{\rm{cl}}$ makes the right hand
side positive, and $\theta$ starts to increase, which eventually causes
the next stick of the CL. However, if the evaporation rate $k_{\rm{ev}}$ is large,
the change of $k_{\rm{cl}}$ does not cause the sign change of
$\dot \theta$, and
$\theta$ keeps decreasing even in the slip state. If this happens,
the contact line keeps receding until it reaches the center, and no ring appears.
Whether the contact angle starts to increase or not when slip starts (i.e.,
when $\theta$ and $k_{\rm{cl}}$ become equal to $\theta_{\rm{R}}$
and 0, respectively) depends on the sign of the right hand side of
Eq.~(\ref{eqn:dangl}): if the sign is negative, the contact angle
keeps decreasing, and no second ring appears, while if the sign
is positive, the contact angle starts to increase, and will
form next inner ring. The condition that multi-ring is observed is
therefore given by
\begin{equation}\label{eqn:sta}
\theta^2_{\rm{R}}\left(\theta^2_e-\theta^2_{\rm{R}}\right)
   \ge
 C \left(\frac{ \theta_0 }{ \sqrt{2\pi} }\right)^{\frac{2}{3}}
    \theta^3_e k_{\rm{ev}},
\end{equation}
where we have set $R$ equal to $R_0$ and $k_{\rm{cl}}=0$.
Equation (\ref{eqn:sta}) gives the
condition for the multi-ring pattern to be observed  for a droplet having
initial radius $R_0$ and initial contact angle $\theta_0$
(Notice that the condition depends on $R_0$ since $k_{\rm{ev}}$ depends on $R_0$).

\begin{figure}[hb]
\begin{center}
\includegraphics[bb=0 0 360 280, scale=0.6,draft=false]{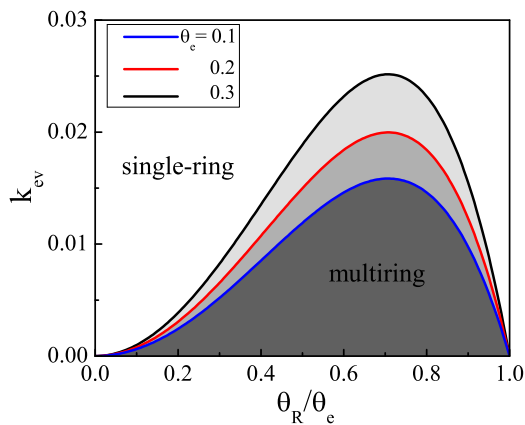}
\caption{
Phase diagram for the multi-ring formation in the plane of the rescaled receding contact angle,
$\theta_{\rm{R}}/\theta_{e}$, and the evaporation rate, $k_{\rm{ev}}$. The lines indicate the boundary between a multi-ring phase below and a single-ring phase above.}
\label{fig:p1}
\end{center}
\end{figure}

Figure~\ref{fig:p1} shows the region defined by Eq.~(\ref{eqn:sta}). Here
$\theta_0$ has been set to be equal to $\theta_e$.
A given droplet can form multi-ring pattern as long as its initial
state is located in the region of "multi-ring". The boundary is
determined by the evaporation rate (characterized by $k_{\rm{ev}}$) and the
three contact angles (equilibrium angle, initial angle and
receding angle). The figure indicates that multi-ring is not observed if the
evaporation rate is large. It also shows that droplets with larger
equilibrium contact angles have larger parameter space for
multi-ring pattern. It should be noted that in this graph the
mountain-like pattern is included in the category of "single-ring"
since there is no inner ring in the mountain-like pattern.

\subsection*{Terminal Behavior of Multi-Ring} \label{sec.3.3}

Figure~\ref{fig:pattern} also shows that even if a droplet leaves many rings as it evaporates, the
ring-pattern disappears in the final stage of drying.
This is consistent with experimental observations~\cite{Adachi95,Chang08,Shmu02,Bi12,Yang14,Jing16}
that the multi-ring pattern is usually made of a solid circle in the center
surrounded by many concentric rings.
Here we discuss how the ring pattern changes to solid circle pattern, and
estimate the radius of the innermost ring $R_{\rm{S}}$.

\begin{figure}[h!]
\begin{center}
\includegraphics[bb=0 0 360 170, scale=1.25,draft=false]{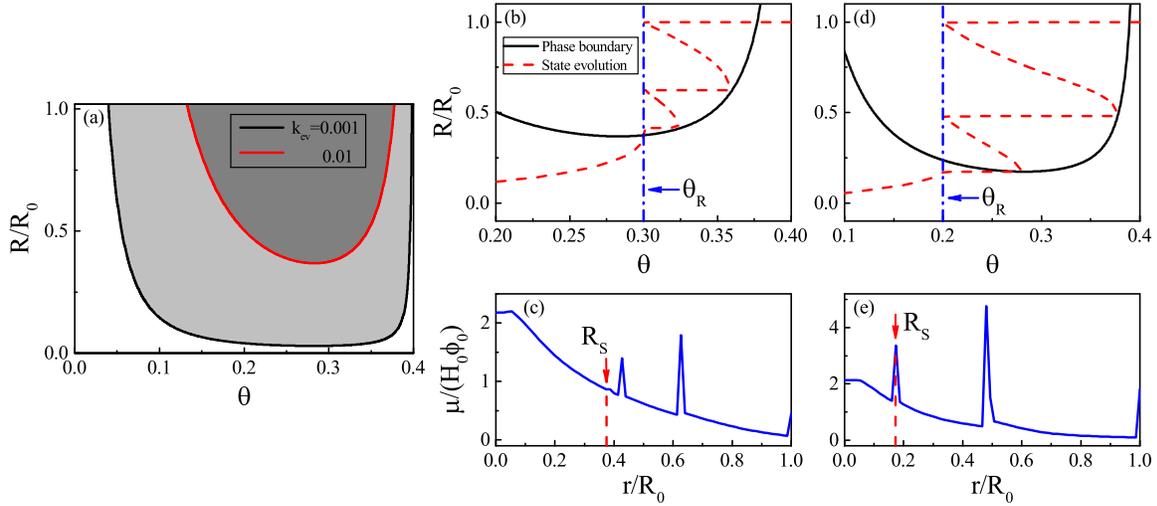}
\caption{Phase space of the dynamical model for evaporating droplet. The dark region shows the $\dot \theta >0$ region for a droplet in depinned state. If a depinned droplet described by ($\theta$, $R/R_{\rm{0}}$) is located in this region, it will be pinned and form a ring. On the other hand, if the droplet is outside of this region, it will form a solid-circle instead. For Figs. (b) to (e): in both up column figures, the red dashed lines are the evolving trajectories of $\theta(t)$ and $R(t)$ of droplet, while the black solid lines denotes the boundary of $\dot \theta =0$; Both down column figures are the corresponding deposition patterns. The arrows in Figs.~(c) and (e) indicate the positions at which the multi-ring terminates. The calculation parameters are: $k_{\rm{ev}}=0.01$ and $\theta_{\rm{R}}=0.3$ in (b) and (c), and $k_{\rm{ev}}=0.005$ and $\theta_{\rm{R}}=0.2$ in (d) and (e), while $\theta_e=\theta_0=0.4$ for all figures.}
\label{fig:p2}
\end{center}
\end{figure}

We consider a 2D parameter space made of $\theta$ and $R$, and investigate the time
evolution of the droplet state point $(\theta, R)$ during evaporation in this space.
The contact line is in
the stick state when $k_{\rm{cl}}$ is $\alpha$
(a large positive value), and in the slip state when $k_{\rm{cl}}$ is zero.
The transition from the stick state to the slip state
takes place when
\begin{equation}\label{eqn:dyn2}
   \theta=\theta_{\rm{R}}.
\end{equation}

On the other hand, the transition from the slip state to the stick state takes place
when $\dot \theta $ given by Eq.~(\ref{eqn:dangl}) becomes zero, or $R$ is equal to $g(\theta)$ defined by
\begin{equation}\label{eqn:dyn}
g(\theta) = \frac{2C\theta^3_e V^{\frac{2}{3}}_{0}}{\pi R_0} \frac{k_{\rm{ev}}}{\theta^2\left(\theta^2_e-\theta^2\right)}.
\end{equation}

The lines $R=g(\theta)$
are shown in Fig.~\ref{fig:p2}a for two evaporation rates, $k_{\rm{ev}}=10^{-2}$
and $k_{\rm{ev}}=10^{-3}$. The line separates the $\dot \theta >0$ region above
and $\dot \theta <0$ region below. This phase diagram shows that
when the droplet contact line is in the slip state: if the state point
$(\theta, R)$ is in the $\dot \theta >0$ region
the contact angle increases until it arrives at the boundary line, where the slip-to-stick
transition takes place resulting in the multi-ring pattern;
If the state point is in the $\dot \theta <0$ region,
the contact line can not stick again resulting in the solid-circle pattern.
It is shown that the range of parameter space for
multi-ring formation is wider for slower evaporation ($k_{\rm{ev}}=10^{-3}$)
than faster evaporation ($k_{\rm{ev}}=10^{-2}$).

These two lines $\theta=\theta_{\rm{R}}$ and $R=g(\theta)$ in the $(\theta, R)$ space define a region
$\theta >\theta_{\rm{R}}$ and $R>g(\theta)$, which we shall call the oscillatory region.
If the state point $(\theta, R)$ is in the oscillatory region, the point
goes back and forth between the two lines and the
stick-slip motion occurs, as shown in Figs.~\ref{fig:p2}b and \ref{fig:p2}d by the red-dashed lines.
In this case, multi-ring
deposition pattern appears, where each ring corresponds to a
stick state. If the state point $(\theta, R)$ moves out of the region,
the stick-slip motion terminates leading to the termination of ring-pattern.

There are two ways for the state point to move out of the
oscillatory region. Let $\theta_c$ be the angle at which $g(\theta)$ becomes minimum.
Using Eq.~(\ref{eqn:dyn}), we have $\theta_c=\theta_e/\sqrt{2}$.
If $\theta_R > \theta_c$, the state point moves out of the
oscillatory region passing through the intersection of the two lines
$\theta=\theta_R$ and $R=g(\theta)$ (see Fig. \ref{fig:p2}b). In this case,
the radius $R_{\rm{S}}$ of the innermost ring is given by the $R$ coordinate of the intersection, i.e.,
$R_{\rm{S}}=g(\theta_R)$, or by use of Eq.~(\ref{eqn:dyn})
\begin{equation}\label{eqn:rs}
R_{\rm{S}} = \frac{2C\theta^3_e V^{\frac{2}{3}}_{0}}{\pi R_0} \frac{k_{\rm{ev}}}{\theta_{\rm{R}}^2\left(\theta^2_e-\theta_{\rm{R}}^2 \right)}.
\end{equation}
On the other hand, if $\theta_{\rm{R}} < \theta_c$ the state point can be outside
of the oscillatory region by moving out of the region $R>g(\theta)$
(see Fig. \ref{fig:p2}d). In this case, $R_{\rm{S}}$ is given by the radius of the last
stick state, and depends on the initial state. We can use the
minimum of the curve $R=g(\theta)$ to estimate $R_{\rm{S}}$, ie.,
$R_{\rm{S}} \approx g(\theta_c)$. Using Eq.~(\ref{eqn:dyn}), we have
\begin{equation}\label{eqn:cri}
 R_{\rm{S}}=\frac{8C V^{\frac{2}{3}}_{0} k_{\rm{ev}}} {\pi R_0 \theta_e}.
\end{equation}

These theoretical values of $R_{\rm{S}}$ are shown by arrows in Figs.~\ref{fig:p2}c and \ref{fig:p2}e,
which are the deposition patterns of Fig.~\ref{fig:p2}b and \ref{fig:p2}d, respectively.
It is seen that they represent well the terminal point of the multi-ring
pattern.

\section*{Conclusion}

In this paper, we have proposed a simple model for the formation of the multi-ring pattern that is often observed in drying droplets. The model predicts that the multi-ring pattern appears only when the evaporation rate is less than a certain critical value determined by the equilibrium contact angle $\theta_e$ and the receding contact angle $\theta_{\rm{R}}$. The model shows that as the ring radius decreases and below a critical value (the innermost ring radius), the multi-ring is replaced by a solid-circle pattern. Analytical expressions have been given for the radius of the innermost ring in terms of $\theta_e$, $\theta_{\rm{R}}$ and the evaporation rate. These results agree with existing experiments qualitatively, and can be tested quantitatively. The remaining question is how the key parameters in the model changes with the solute details (concentration~\cite{Chang08,Orejon11}, particle size~\cite{Shmu02,Zhang08} and shape~\cite{Peter11,Alex15} etc), and such effects will be discussed in future.

\bigskip
{\bf Acknowledgement.}~~
This work was supported in part by Grant No. 21434001 and 21404003
of the National Natural Science Foundation of China (NSFC), the joint NSFC-ISF Research Program, jointly funded by the
NSFC under Grant No. 51561145002 and the Israel
Science Foundation (ISF) under Grant No. 885/15, and the Fundamental Research Funds for the Central Universities.


\newpage
\vskip 0.5truecm
\centerline{for Table of Contents use only}
\centerline{\bf Multi-Ring Deposition Pattern of Drying Droplets}
\centerline{\it Mengmeng Wu, Xingkun Man, and Masao Doi}

\begin{figure}[h]
\begin{center}
\includegraphics[bb=0 0 536 370, scale=0.5,draft=false]{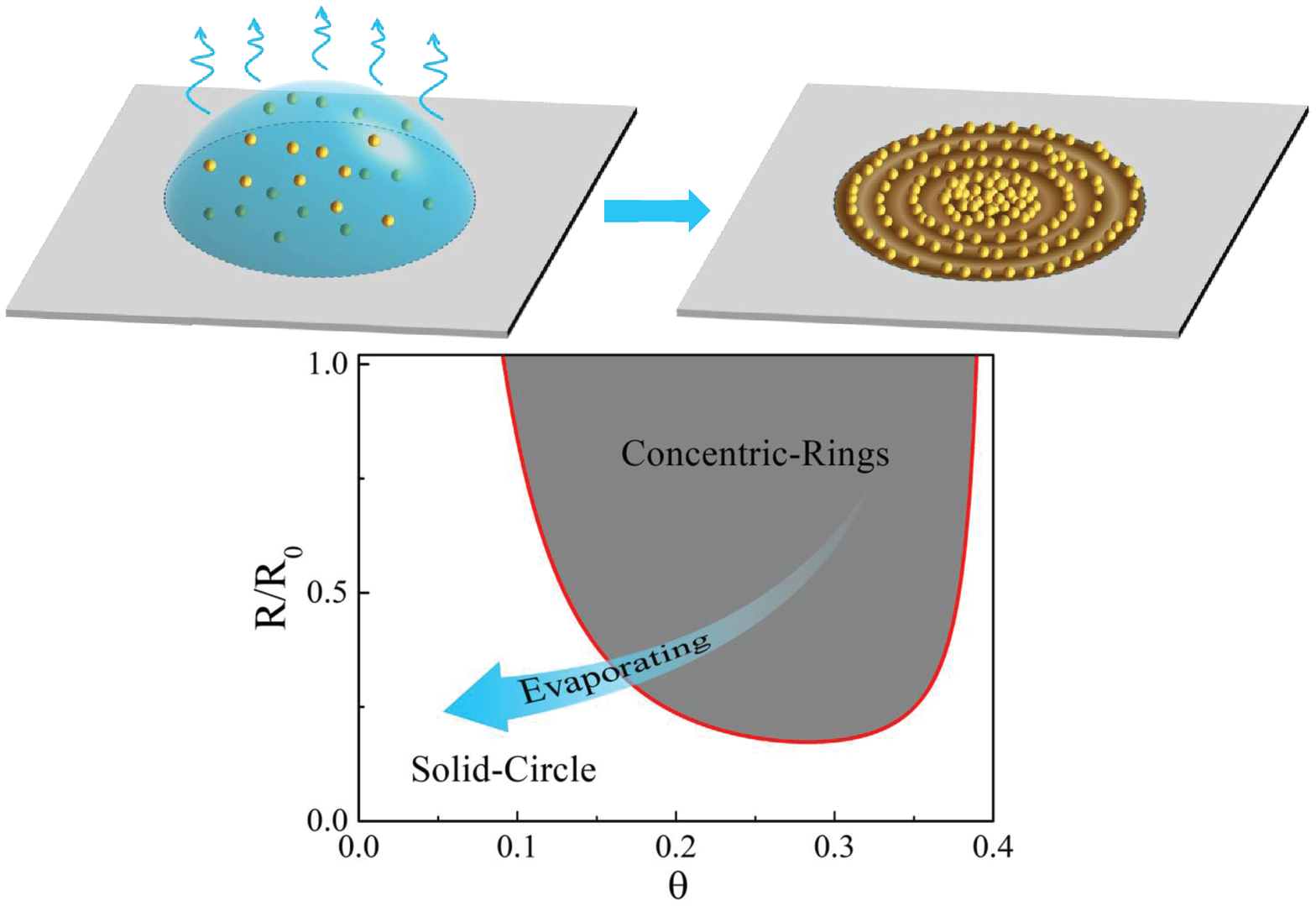}
\end{center}
\end{figure}

\end{document}